\documentclass[final,twocolumn]{IEEEtran}

\usepackage{verbatim}
\usepackage{rotating, booktabs}  
\usepackage{subcaption}
\usepackage{graphics}
\usepackage{graphicx}
\usepackage{amsthm}
\usepackage{braket}
\usepackage{hyperref}
\usepackage{listings}
\usepackage{color, xcolor}
\usepackage{upgreek}
\usepackage{dcolumn}
\newcolumntype{z}[1]{D{.}{.}{#1}}
\usepackage{float} 
\usepackage{algorithm}
\usepackage[noend]{algpseudocode}
\usepackage{extpfeil}
\usepackage{etoolbox}
\usepackage{url}
\usepackage{multirow} 
\usepackage[mathscr]{eucal}
\usepackage{shadow,epsf,amsthm,amssymb,amsmath}
\usepackage{algorithm}
\usepackage{algpseudocode}
\usepackage[switch]{lineno}

\newtheorem{definitionenv}{Definition}
\newtheorem{lemmaenv}[definitionenv]{Lemma}
\newtheorem{theoremenv}[definitionenv]{Theorem}
\newtheorem{corollaryenv}[definitionenv]{Corollary}
\newtheorem{propositionenv}[definitionenv]{Proposition}
\newtheorem{remarkenv}[definitionenv]{Remark}
\newenvironment{remark}{\begin{remarkenv}\rm}{\end{remarkenv}}
\newcommand{\br}{\begin{remark}}
\newcommand{\er}{\end{remark}}

\newtheorem{exampleenv}{Example}
\newtheorem{app-lemmaenv}[section]{Lemma}

\newenvironment{definition}{\begin{definitionenv}\rm}{\end{definitionenv}}
\newenvironment{lemma}{\begin{lemmaenv}\rm}{\end{lemmaenv}}
\newenvironment{theorem}{\begin{theoremenv}\rm}{\end{theoremenv}}
\newenvironment{corollary}{\begin{corollaryenv}\rm}{\end{corollaryenv}}
\newenvironment{example}{\begin{exampleenv}\rm}{\end{exampleenv}}
\newenvironment{proposition}{\begin{propositionenv}\rm}{\end{propositionenv}}
\newenvironment{app-lemma}{\begin{app-lemmaenv}\rm}{\end{app-lemmaenv}}

\newcommand{\bd}{\begin{definition}}
	\newcommand{\ed}{\end{definition}}
\newcommand{\bl}{\begin{lemma}}
	\newcommand{\el}{\end{lemma}}
\newcommand{\elp}{\hspace*{\fill} $\Box$
\end{lemma}}
\newcommand{\bt}{\begin{theorem}}
\newcommand{\et}{\end{theorem}}
\newcommand{\etp}{\hspace*{\fill} $\Box$
\end{theorem}}
\newcommand{\bc}{\begin{corollary}}
\newcommand{\ec}{\end{corollary}}
\newcommand{\ecp}{\hspace*{\fill} $\Box$
\end{corollary}}

\newcommand{\be}{\begin{example}}
\newcommand{\ee}{\end{example}}
\newcommand{\eep}{\hspace*{\fill} $\Box$
\end{example}}
\newcommand{\bp}{\begin{proposition}}
\newcommand{\ep}{\end{proposition}}
\newcommand{\epp}{
\end{proposition}}

\usepackage{xspace}

\usepackage[colorinlistoftodos,prependcaption]{todonotes}
\usepackage{xargs}

\newcommandx{\yellownote}[2][1=]{\todo[inline,linecolor=yellow,backgroundcolor=yellow!25,bordercolor=yellow,#1]{#2}}
\newcommandx{\greennote}[2][1=]{\todo[inline,linecolor=green,backgroundcolor=green!25,bordercolor=green,#1]{#2}}

\newcommand{\cT}{{\cal T}}

\allowdisplaybreaks

\begin{document}

\title{Loss-Tolerant Measurement-Device-Independent QKD over Untrusted Repeaters via High-Dimensional Time-Bin Multiplexing}

\author{Yu-Chun~Yin,  Ching-Yi~Lai, \,and\, Alexei Ashikhmin
\thanks{\footnotesize
YCY and CYL are with the Institute of Communications Engineering, National Yang Ming Chiao Tung University, Hsinchu 300093, Taiwan. (email: cylai@nycu.edu.tw)}
\thanks{\footnotesize
AA is with Nokia Bell Labs, Murray Hill, New Jersey, USA.}
\thanks{\footnotesize
CYL acknowledges support from the National Science and Technology Council in Taiwan under Grants 113-2221-E-A49-114-MY3  and 114-2119-M-A49-006-.}
}
\maketitle

\begin{abstract}

We propose a loss-tolerant measurement–device-independent quantum key distribution (MDI-QKD) protocol using time-bin–multiplexed single photons entangled with local qubits stored in nitrogen-vacancy (NV) centers. In each round, Alice and Bob each send a single multiplexed photon to an untrusted intermediary, Charlie. Each photon first becomes entangled with Charlie’s NV-center qubits; Charlie then performs entanglement swapping between his NV qubits to establish EPR pairs between Alice’s and Bob’s local qubits. With $d$ time bins per photon, $\log_2 d + 1$ EPR pairs can be generated per transmitted photon, substantially increasing the probability of EPR-pair creation under limited channel bandwidth. 
To further mitigate the effects of prolonged memory storage, the protocol employs a memory-insensitive (MI) architecture, where multiple banks of qubits and multiplexed photon routing reduce the effective memory lifetime requirements. This approach enhances loss tolerance, enabling higher secret key rates and secure key distribution over longer distances compared to conventional single-photon protocols. Using the Lo–Chau EPR–QKD framework and Shor–Preskill reduction, we prove that the protocol is secure without trusting intermediate nodes, making it suitable for quantum networks with untrusted repeaters.

\end{abstract}

\date{\today}

\section{Introduction and Motivation}

Future quantum communication networks~\cite{AZ2021,ML2016} promise powerful capabilities,
including long-distance entanglement distribution~\cite{Gisin2007,Ursin2007},
distributed quantum computation~\cite{CEHM1999,SMB2006,LTSL2007},
and delegated quantum information processing~\cite{Fitzsimons2017}.

Quantum key distribution (QKD) enables provably secure classical communication based on the principles of quantum mechanics.
Measurement-device-independent (MDI) QKD protocols~\cite{LCQ12,BP12} further enhance security by removing vulnerabilities associated with imperfect measurement devices.
Among different physical implementations, QKD using high-dimensional encoding offers the advantage of increased key rates~\cite{DSB18,NMW2017,CLK2022},~\cite{Chang2024,LH2024}.

Recent work has shown that high-dimensional time-bin–multiplexed photons can reduce the number of transmitted photons and other resources, or improve fidelity under channel loss~\cite{LMN19,XLM+21,PHG+20}. 
In such schemes, a single photon can be coherently delocalized across multiple time bins, each serving as a distinct temporal mode.
Each temporal mode can then be entangled with a local physical qubit stored in a quantum memory (QM) in a nitrogen-vacancy (NV) center, creating a superposition of time-labeled EPR pairs. 
Such high-dimensional time-bin encoding has also been experimentally demonstrated in the implementation of a compiled version of Shor’s factoring algorithm using a single photon~\cite{WC2024}.

In this paper, we propose QKD protocols that employ time-bin multiplexing of single photons, offering improved key rates and extended transmission distances. In particular, we present a measurement-device-independent QKD (MDI-QKD) scheme in which intermediate nodes are untrusted. 
In a simple scenario, Alice and Bob each generate EPR pairs with an intermediary, Charlie, who then performs entanglement swapping to establish EPR pairs directly between Alice and Bob~\cite{BBM92,LC1999,BDCZ1998,AT2023}. 
With $d$ time bins per photon, $\log_2 d+1$ EPR pairs can be generated by transmitting one photon from each of Alice and Bob, making repeater nodes significantly more effective in extending the reach of QKD~\cite{BDCZ98,MT13}. 
As a result, our protocols increase the secret-key rate of memory-assisted (MA) QKD~\cite{NMW2017,LRM2017} and its MDI variant by a factor that scales as $\log_2 d$, yielding a logarithmic improvement over conventional schemes.

To evaluate the performance of the proposed scheme, we derive the average secret key rate of our MA time-bin multiplexed QKD protocol using the Shor--Preskill approach~\cite{SP00} with realistic experimental parameters. We further determine the optimal depth of time-bin multiplexing for two parties separated by $300$ km. Our results show that time-bin multiplexed QKD protocols significantly outperform conventional ones under bandwidth constraints. Without multiplexed photons, conventional schemes require longer quantum memory storage to accumulate sufficient EPR pairs, which amplifies errors in imperfect memories. For example, at a secret key rate of $10^{-4}$, our multiplexed protocol with $\log_2 d = 7$ achieves a maximum distance of $217.8$ km, compared to less than $133$ km for the conventional case.

Finally, we propose an entanglement-based MDI-QKD protocol that incorporates time-bin multiplexing with an untrusted intermediate node. By employing a memory-insensitive (MI) technique~\cite{CK2007}, we show that multi-banks of QMs together with multiplexed photon routing can effectively mitigate the prolonged storage of entanglement in quantum memories. As a result, the effective transmission distance at which the key rate drops to $90\%$ of its maximum nearly doubles, increasing from approximately $169-173$ km without a repeater to around $324-331$ km with a repeater.

\begin{table*}[htbp]
\centering
\caption{Comparison of high-dimensional QKD protocols.}
\label{tb:comparison_hdQKD}
\begin{tabular}{p{3.cm} p{4.5cm} p{4.5cm} p{4.5cm}}
\hline
\textbf{Feature} & \textbf{This Work (MA-MDI-QKD)} & \textbf{Dellantonio et al. \cite{DSB18}} & \textbf{Chang et al. \cite{Chang2024}} \\
\hline
QKD Type & Entanglement-based & Prepare-and-measure & Entanglement-based \\
\hline
Security & MDI & MDI & One-sided DI (Steering-based) \\
\hline
Quantum Memory & \textbf{Yes} (NV-center) & No & No \\
\hline
High-Dimension Source & Multiplexed generation of EPR pairs between QMs via time-bin encoding & High-dimensional encoding in spatial/temporal modes & Large-alphabet time-bin encoding\\
\hline
Repeater Capability & \textbf{Yes} (Supports entanglement swapping) & No (Direct transmission) & No (Direct transmission) \\
\hline
PNS Attack Immunity & Inherently immune (Entanglement) & Not inherent (Needs decoy states) & Inherently immune (Entanglement) \\
\hline
Complexity & High (QM + $n$-MUX operations) & Medium (Bell Measurement) & Medium (Dispersive optics sync) \\
\hline
\end{tabular}
\end{table*}

 \begin{figure}
    \centering
    \begin{subfigure}[H]{0.99\linewidth}
        \centering
        \includegraphics[width=0.8\linewidth]{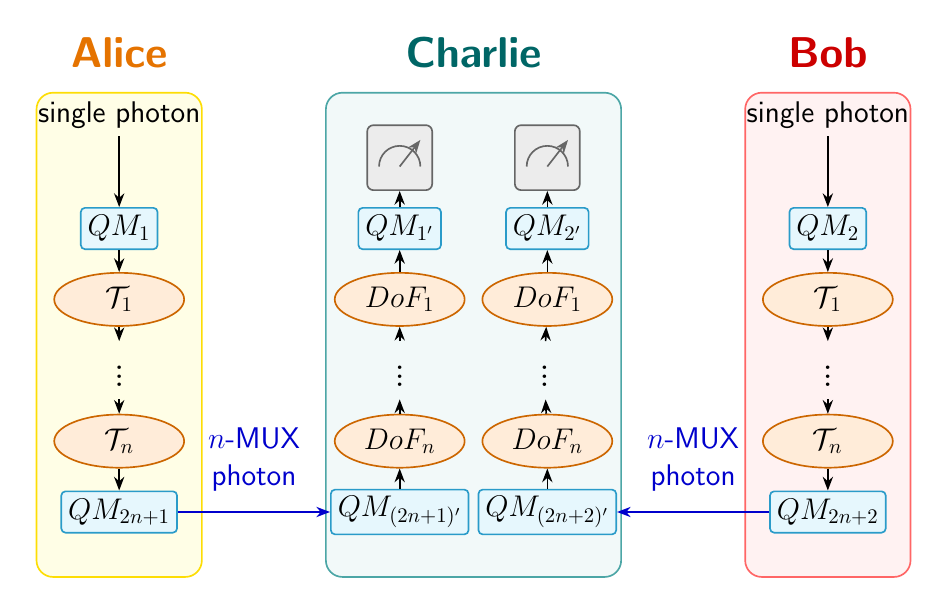}
        \caption{Preparation and transmission}
     \end{subfigure} 
     \begin{subfigure}[H]{0.99\linewidth}
        \centering
        \includegraphics[width=0.8\linewidth]{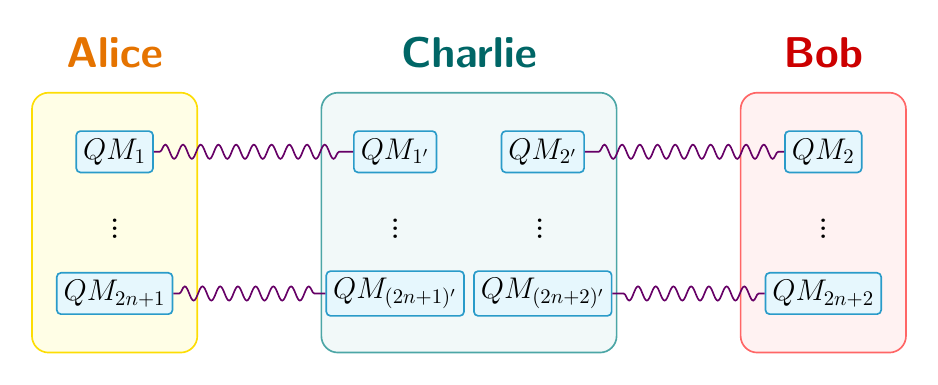}
        \caption{Link entanglement generation}
     \end{subfigure}
      \begin{subfigure}[H]{0.99\linewidth}
        \centering
        \includegraphics[width=0.8\linewidth]{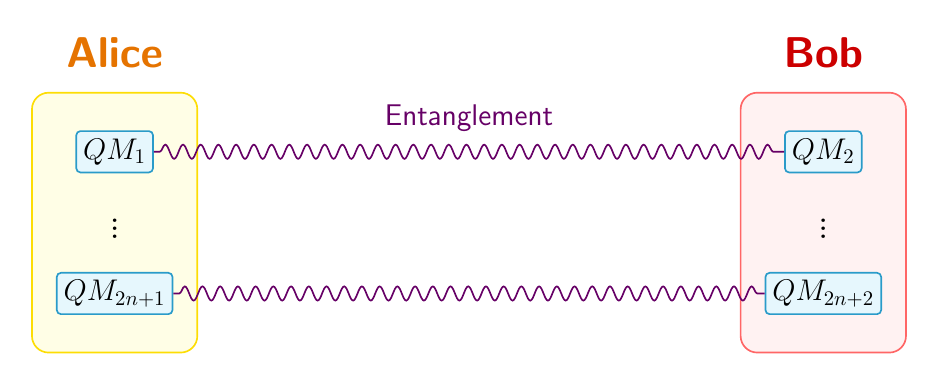}
        \caption{Entanglement swapping}
     \end{subfigure}

\caption{Three-stage operation of the $n$-MUX quantum repeater scheme. (a) Alice and Bob generate $n$-MUX photons and send them to the intermediate node, Charlie. (b) Successful optical measurements at Charlie herald the creation of entanglement across the elementary links (A–C and B–C). (c) Charlie subsequently performs entanglement swapping on his QMs, thereby distributing end-to-end entanglement between Alice and Bob for subsequent QKD operations.}
         \label{fig:1}
\end{figure}

Unlike previous MA-MDI-QKD schemes~\cite{LRM2017_2, LRM2017_11}, which require asynchronous state loading through teleportation and therefore hinder their application in standard entanglement-swapping-based repeaters, our protocol adopts an end-to-end entanglement-based approach. This design enables efficient sequential repeater operations for robust long-distance secure communication.

In contrast, while~\cite{DSB18,Chang2024} employ high-dimensional spatial or temporal encoding to maximize the information carried by a single photon, our architecture uses a single multiplexed photon to interact with multiple local quantum memories simultaneously. This enables the one-shot generation of multiple EPR pairs between Alice’s and Bob’s QMs, providing a high-capacity entanglement resource that is fundamentally different from simple qudit encoding.

Furthermore, this entanglement-based design is inherently immune to photon-number-splitting (PNS) attacks~\cite{ZH2022}. By contrast, the high-dimensional MDI-QKD scheme in~\cite{DSB18} follows a prepare-and-measure paradigm and therefore lacks intrinsic immunity to PNS attacks.

Although~\cite{Chang2024} also considers entanglement distribution, their protocol establishes security through the steering probabilities of time-bin measurement outcomes. In contrast, our protocol uses a single multiplexed photon to directly generate multiple EPR pairs between Alice’s and Bob’s quantum memories, which are subsequently distilled for secure key generation.

Table~\ref{tb:comparison_hdQKD} summarizes these distinctions.

Overall, unlike previous MA-MDI-QKD schemes~\cite{LRM2017_2, LRM2017_11}, which require additional entanglement infrastructure to support repeater operations, our entanglement-based approach provides native repeater compatibility. EPR generation, swapping, and distillation are intrinsic components of the protocol rather than external architectural extensions. It is important to note that the $n$-MUX entanglement generation process introduces higher preparation complexity and an initial rate trade-off compared to conventional prepare-and-measure schemes, which rely on simpler single-photon state preparation.

 This approach integrates the efficiency of mode multiplexing with the robustness of quantum error correction, opening a new direction for fault-tolerant QKD in excitation-conserving photonic systems. 
 Looking ahead, such approaches may contribute to the long-term vision of a fault-tolerant quantum internet.
In such a network, error-corrected logical qubits encoded across multiple entangled physical qubits can be transmitted to enable scalable and reliable global quantum communication~\cite{LK24}. In this context, quantum-multiplexed CSS QKD protocols based on time-bin multiplexing of photons can be further developed~\cite{PHG+20}. 

The MATLAB source code for the QKD key rate simulations and the performance analysis presented in this paper is available at:~\cite{YCY2026}.

This paper is organized as follows.
In the next section, we introduce the notation and the concept of heralded entanglement generation using quantum multiplexed photons.
Section~\ref{sec:mux_QKD} presents our MA time-bin multiplexed QKD protocol, including its key rate analysis and a comparison with conventional protocols.
In Section~\ref{sec:mux_MDIQKD}, we extend the discussion to MDI-QKD protocols.
Finally, Section~\ref{sec:con} concludes the paper.

 Section~\ref{sec:mux_QKD} established the theoretical 
foundation using a single memory bank, demonstrating how multiplexing 
improves secret key rates under bandwidth constraints.

Building on this foundation, we proposed a practical MDI-QKD protocol 
(Section~\ref{sec:mux_MDIQKD}) incorporating multiple memory banks 
for pipelined operation. This memory-insensitive architecture represents 
a realistic implementation suitable for quantum networks with untrusted 
repeaters, achieving substantially higher throughput than sequential 
single-bank schemes.

\section{Entanglement Generation Using Time-Bin Multiplexing of Photons}

In this section, we introduce the polarization encoding of single photons combined with time-bin multiplexing, as well as their interactions with qubits stored in quantum memories. The formalism presented here is general and does not depend on a specific physical implementation of QM. For concreteness, we use NV-center-based QMs, following the implementations described in~\cite{LMN19,XLM+21}.
Each QM is modeled as a two-level system with basis states 
$\ket{\mathrm{0}}$  (ground) and $\ket{\mathrm{1}}$ (excited),
and we denote the state of memory $i$ as $\ket{b}_i$. 
In a two-party protocol with Alice and Bob, QMs with odd indices are assigned to Alice, while those with even indices are assigned to Bob.

For simplicity, we describe these operations using the language of the quantum circuit model in the computational basis, without detailing the underlying optical components. This approach allows readers unfamiliar with specific optical instruments to understand the operational meaning of the protocol, while the corresponding experimental realizations can be found in~\cite{LMN19,XLM+21}.

\subsection{Time-Bin Multiplexing of Photons}

A single photon can be polarized in the diagonal or anti-diagonal direction, denoted by $\ket{D}$ or $\ket{A}$. 
In an $n$-multiplexed photon (abbreviated $n$-MUX), where $n \ge 0$, the photon’s time duration is divided into $2^{n}$ equal-sized time bins, indexed by $n$ bits. 
The photon thus carries $n$ time-bin bits plus one polarization bit, giving a total of $n+1$ classical bits per photon. 
In the computational basis, we define
\begin{align}
\ket{0}_{P} \equiv \ket{D}, \quad \ket{1}_{P} \equiv \ket{A},
\end{align}
so that an $n$-MUX photon can be expressed as
\begin{align}
\ket{b_P}_{P}\otimes \ket{b_1}_{t_1} \cdots \ket{b_n}_{t_n},
\end{align}
where $b_P$ is the polarization bit, and $b_1\cdots b_n$ specify the time-bin index ($0 \le b_1\cdots b_n < 2^n$).
Here, we refer to $n$ as the \emph{multiplexing depth}, corresponding to a total encoding dimension of $2^{n+1}$.

For example, a $2$-MUX photon employs four time bins labeled by $b_1b_2\in\{ 00,01,10,11\}$. In the special case  $n=0$, there is only the polarization degree of freedom with no time-bin multiplexing (0-MUX), so the state reduces to $\ket{b_P}_{P}$.

\subsection{Operations on Multiplexed Photons}

We now define the key quantum operations used in our protocol.

\subsubsection{Delay Operator}

A delay operator $\mathcal{T}_i$ acts on a photon as follows:
\begin{align}
   &(\alpha\ket{0}_{P}+\beta \ket{1}_{P})\ket{0}_{t_1}\cdots\ket{0}_{t_{n}} \notag\\
   \xrightarrow{\cT_i}& \alpha\ket{0}_{P}\ket{0}_{t_1}\cdots\ket{0}_{t_{n}}+\beta \ket{0}_{P}\ket{0}_{t_1}\cdots\ket{1}_{t_i}\cdots\ket{0}_{t_{n}}.
\end{align}

In other words, a photon in state $\ket{D}$ is unaffected by $\cT_i$, while a photon in state $\ket{A}$ is converted to $\ket{D}$ and transferred to the time bin indexed by $i$. In this sense, $\cT_i$ effectively acts as a SWAP operation between the polarization mode and the $i$-th time-bin mode.

\subsubsection{Degree-of-Freedom Swap Gate}
The $SWAP(P,\mathcal{T}_i)$ gate exchanges the polarization bit with 
the $i$-th bit of the time-bin index:
\begin{align}
\ket{b_P}_P \ket{b_i}_{t_i} 
\xrightarrow{SWAP(P,\mathcal{T}_i)}
\ket{b_i}_P \ket{b_P}_{t_i}.
\end{align}
This swap exchanges information stored in polarization and in the $i$-th time-bin bit, enabling flexible routing and reuse of the polarization channel for sequential entanglement generation. 

\subsubsection{Controlled-NOT Gate}
The QM-photon interaction of interest is a QM-controlled polarization-flip gate, denoted  $CNOT_{i,P}$, 
where the state of qubit $i$ controls the flip of a photon's polarization~\cite{LRM2017},~\cite{FB2024},~\cite{CW2025}. 

The $CNOT_{i,P}$ gate flips the photon's polarization conditional on 
the state of qubit $i$:
\begin{equation} \label{MP_intrac}
\begin{aligned}
    \ket{0}_{i}\ket{b}_{P} &\xrightarrow{CNOT_{i,P}} \ket{0}_{i}\ket{b}_{P},\\
    \ket{1}_{i}\ket{b}_{P} &\xrightarrow{CNOT_{i,P}} \ket{1}_{i}\ket{\bar{b}}_{P},
\end{aligned}
\end{equation}
where $\bar{b} = 1-b$.

\subsection{Heralded Entanglement Sharing between QMs using Photon Transmission}

We now describe how two distant parties, Alice and Bob, can establish a Bell state between their QMs using a single transmitted photon in a \emph{heralded} protocol~\cite{LMN19}.

The protocol introduced in~\cite{LMN19} proceeds as follows:
\begin{enumerate}
    \item {Initialization:} Alice and Bob each prepare a QM in the superposition state
    \[
        \ket{+} = \frac{1}{\sqrt{2}}(\ket{0}+\ket{1}),
    \]
    labeled $\ket{+}_1$ (Alice) and $\ket{+}_2$ (Bob).
 Alice also prepares a single photon in the polarization state $\ket{0}_{P}$.

 \item {Photon-QM interaction:}  
   Alice let the photon interact with her QM via $CNOT_{1,P}$ as defined in (\ref{MP_intrac}). The photon is then sent to Bob, who applies $CNOT_{2,P}$ between his QM and the photon. The combined system evolves as:
    \begin{equation}\label{heralded}
       \begin{aligned}
       \ket{+}_{1}\ket{+}_{2}\ket{0}_{P} 
       &\xrightarrow{CNOT_{1,P}} \frac{\ket{0}_{1}\ket{+}_{2}\ket{0}_{P} + \ket{1}_{1}\ket{+}_{2}\ket{1}_{P}}{\sqrt{2}} \\
       &\xrightarrow{CNOT_{2,P}} \frac{\ket{\Phi^+}_{12}\ket{0}_{P} + \ket{\Psi^+}_{12}\ket{1}_{P}}{\sqrt{2}},
       \end{aligned}
    \end{equation}
    where
    $
    \ket{\Phi^+} = \frac{1}{\sqrt{2}}(\ket{00} + \ket{11})$ and $
    \ket{\Psi^+} = \frac{1}{\sqrt{2}}(\ket{01} + \ket{10})
    $
    are Bell states of QMs $1$ and $2$.

    \item {Heralding:} 
     Bob measures the photon’s polarization. If the outcome corresponds to $\ket{1}_{P}$, then a Pauli $X$-correction is applied to Bob's QM. After this step, Alice and Bob share the Bell state $\ket{\Phi^+}$.
\end{enumerate}

\subsection{Heralded Entanglement Sharing between QMs using Time-Bin Multiplexed Photon Transmission}
\label{sec:mux_EPR}

The entanglement-sharing protocol from the previous subsection can be generalized with time-bin–multiplexed photon transmission to generate multiple EPR pairs between QMs. 
The following protocol~\cite{LMN19,XLM+21} generates $n+1$ EPR pairs between Alice’s and Bob’s QMs using an $n$-MUX photon.

\begin{enumerate}
    \item {Initialization:}  
     Alice and Bob prepare QMs $Q_{2j-1}$ and $Q_{2j}$ in the superposition state $\ket{+}$,
    labeled $\ket{+}_{2j-1}$ (Alice) and $\ket{+}_{2j}$ (Bob),  $j=1,\dots, n, n+1$.
    Alice also prepares an $n$-MUX photon in the state $\ket{0}_{P} \ket{0}_{t_1} \cdots \ket{0}_{t_{n}}$.

    \item 
    Alice applies the following sequence:
\begin{align}
    &CNOT_{1,P},\ \cT_1, 
    CNOT_{3,P},\nonumber\\ 
    &\cT_2,
    \dots,
    CNOT_{2n+1,P}.  \label{eq:CT_interaction}
\end{align}
     
    She then transmits the photon to Bob.

    \item 
    Upon receiving the photon, Bob applies:
    \begin{align}
        &CNOT_{2n+2,P},\ SWAP(P,\cT_{n-1}), \
        CNOT_{2n,P}, \nonumber\\
        &SWAP(P,\cT_{n-2}), \dots, 
        CNOT_{2,P}. \label{eq:CS_interaction} 
        \end{align}

    After these steps, the joint state of the QMs and the photon is
    \begin{align}
    &\bigg(\ket{\Phi^+}_{1,2}\ket{0}_P + \ket{\Psi^+}_{1,2}\ket{1}_P\bigg) \notag\\
    &\bigotimes_{i=2}^{n+1} \bigg(\ket{\Phi^+}_{2i-1,2i}\ket{0}_{t_{i-1}} + \ket{\Psi^+}_{2i-1,2i}\ket{1}_{t_{i-1}}\bigg) 
    \end{align}

    \item   {Heralding:}
Bob measures the photon and gets outcomes $b_P b_1 \dots b_{n}$.  Bob applies Pauli-$X$ corrections conditioned on these outcomes, resulting in $n+1$ shared Bell states $\ket{\Phi^+}$.
\end{enumerate}

The protocol exploits time-bin multiplexing to transform a single photon 
into an effective multi-qubit flying register. Through Alice's sequential 
CNOT operations and delay operators, the photon is transformed into a 
multipartite entangled state with multiple local QMs simultaneously. 
When the photon reaches Bob, his QMs interact with it, and the subsequent 
photon measurement heralds $n+1$ EPR pairs shared between Alice's and Bob's QMs.

As an example, we consider the case $n=1$, where a single 1-MUX photon is used to generate two EPR pairs. We take the register order as QM1, QM3, P, $t_1$, QM2, QM4. The initial state is $\ket{\psi_0} = \ket{+}_1 \ket{+}_3 \ket{0}_P \ket{0}_{t_1} \ket{+}_2 \ket{+}_4.$ Alice performs the following operations:

\begin{align*}
    &\ket{+}_{1}\ket{+}_{3}\ket{0}_{P}\ket{0}_{t_1}\ket{+}_{2}\ket{+}_{4}  \\
    \xrightarrow{CNOT_{1,P}}&\frac{1}{\sqrt{2}}\ket{0}_{1}\ket{+}_{3}\ket{0}_{P}\ket{0}_{t_1}\ket{+}_{2}\ket{+}_{4} \\
    +&\frac{1}{\sqrt{2}}\ket{1}_{1}\ket{+}_{3}\ket{1}_{P}\ket{0}_{t_1}\ket{+}_{2}\ket{+}_{4} \\
    \xrightarrow{\cT_1}&\frac{1}{\sqrt{2}}\ket{0}_{1}\ket{+}_{3}\ket{0}_{P}\ket{0}_{t_1}\ket{+}_{2}\ket{+}_{4} \\
    +&\frac{1}{\sqrt{2}}\ket{1}_{1}\ket{+}_{3}\ket{0}_{P}\ket{1}_{t_1}\ket{+}_{2}\ket{+}_{4}\\\
    \xrightarrow{CNOT_{3,P}}&\frac{1}{\sqrt{2}}\ket{0}_{1}\ket{0}_{3}\ket{0}_{P}\ket{0}_{t_1}\ket{+}_{2}\ket{+}_{4} \\
    +&\frac{1}{\sqrt{2}}\ket{1}_{1}\ket{0}_{3}\ket{0}_{P}\ket{1}_{t_1}\ket{+}_{2}\ket{+}_{4}\\\
    +&\frac{1}{\sqrt{2}}\ket{0}_{1}\ket{1}_{3}\ket{1}_{P}\ket{0}_{t_1}\ket{+}_{2}\ket{+}_{4} \\
    +&\frac{1}{\sqrt{2}}\ket{1}_{1}\ket{1}_{3}\ket{1}_{P}\ket{1}_{t_1}\ket{+}_{2}\ket{+}_{4}
    \end{align*}
The photon is sent to Bob, who then performs the following operations:
    \begin{align*}
    \xrightarrow{CNOT_{4,P}}&
\frac{1}{2}(\ket{\Phi^+}_{34}\ket{0}_{P}+\ket{\Psi^+}_{34}\ket{1}_{P})\ket{0}_{t_1}\ket{0}_{1}\ket{+}_{2} \\
    +&\frac{1}{2}(\ket{\Phi^+}_{34}\ket{0}_{P}+\ket{\Psi^+}_{34}\ket{1}_{P})\ket{1}_{t_1}\ket{1}_{1}\ket{+}_{2}\\
    \xrightarrow{SWAP(P,t_1)}&
\frac{1}{2}(\ket{\Phi^+}_{34}\ket{0}_{1}+\ket{\Psi^+}_{34}\ket{1}_{1})\ket{0}_{P}\ket{0}_{1}\ket{+}_{2} \\
    +&\frac{1}{2}(\ket{\Phi^+}_{34}\ket{0}_{1}+\ket{\Psi^+}_{34}\ket{1}_{1})\ket{1}_{P}\ket{1}_{1}\ket{+}_{2}\\
     \xrightarrow{CNOT_{2,P}}     &\frac{1}{2}\ket{\Phi^+}_{12}\ket{\Phi^+}_{34}\ket{0}_{P}\ket{0}_{t_1}\\
+&\frac{1}{2}\ket{\Phi^+}_{12}\ket{\Psi^+}_{34}\ket{0}_{P}\ket{1}_{t_1}\\
+&\frac{1}{2}\ket{\Psi^+}_{12}\ket{\Phi^+}_{34}\ket{1}_{P}\ket{0}_{t_1}\\
+&\frac{1}{2}\ket{\Psi^+}_{12}\ket{\Psi^+}_{34}\ket{1}_{P}\ket{1}_{t_1}.
\end{align*}

Measuring the photon yields outcomes $b_P b_1 \in \{00,01,10,11\}$.
Apply $X$ to $Q_2$ if $b_P=1$ and to $Q_4$ if $b_1=1$.   
After these local corrections, Alice and Bob share two Bell pairs $\ket{\Phi^+}_{12}$ and $\ket{\Phi^+}_{34}$.

 \subsection{Physical Implementation of Quantum Operations}

 \begin{figure}
    \centering
    \begin{subfigure}[b]{0.99\linewidth}
        \centering
        \includegraphics[width=\linewidth]{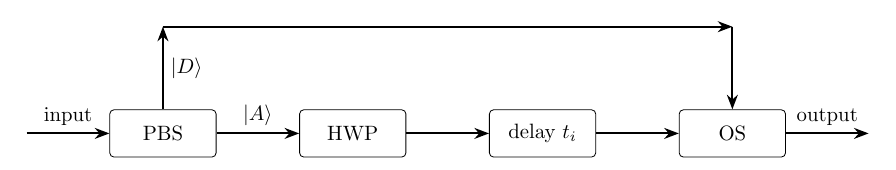}
        \caption{Delay operator $\cT_i$. 
       This circuit delays photons in the $\ket{A}$ polarization by a specific number of time bins while leaving $\ket{D}$ photons unaffected.  
       The delay $t_i$ is chosen according to the delay operator $\cT_i$, which imposes the corresponding time delay.
       }
        \label{fig:delay}
     \end{subfigure}

  \vspace{0.5cm}

    \begin{subfigure}[b]{0.99\linewidth}
        \centering
        \includegraphics[width=\linewidth]{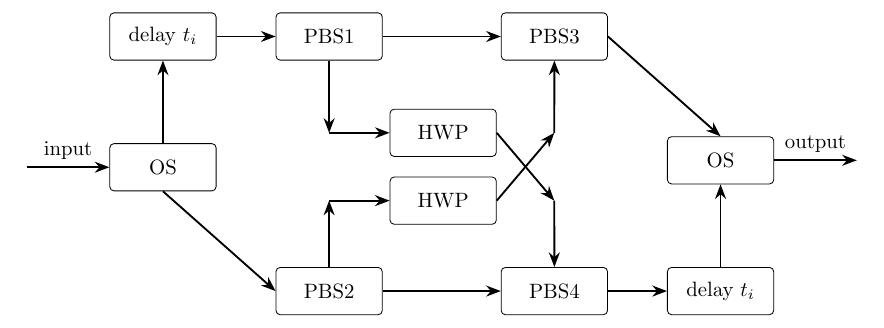}
        \caption{Degree-of-freedom swap gate $SWAP(P,\cT_i)$. This circuit exchanges the polarization bit $b_P$ of a photon with the $i$-th bit $b_i$ of its time-bin index. }

        \label{fig:DOF}
    \end{subfigure}
     
      \vspace{0.5cm}
      
    \begin{subfigure}[b]{0.99\linewidth}
        \centering
        \includegraphics[width=0.8\linewidth]{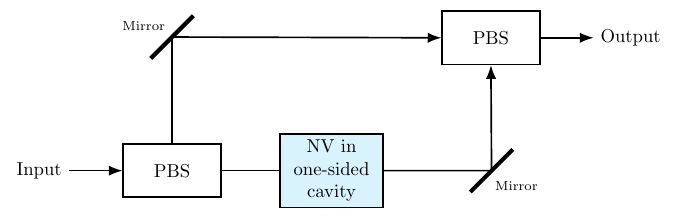}
        \caption{QM-controlled polarization-flip gate. The first PBS splits the input photon into two paths. Upon reflection from the NV-cavity system, a QM-state-dependent phase shift is induced. When these paths recombine at the second PBS, the resulting interference translates the phase shift into a deterministic flip of the photon's polarization state.}
        \label{fig:NV_CNOT}
    \end{subfigure}

    \caption{Optical implementations of (a)  Delay operator, (b) Degree of freedom swap gate, and (c) QM-controlled polarization-flip gate. Components: PBS is polarization beam splitter, HWP is half-wave plate, and OS is optical switch. See~\cite{LMN19} for details of these basic modules.}
    \label{fig:optical}
\end{figure}

Having defined the mathematical operations in the previous section, we now describe their physical realization using optical circuits and NV-center quantum memories, as illustrated in Figure~\ref{fig:optical}.  

1. {$\mathcal{T}_i$:} Photons are separated by polarization using a polarizing beam splitter (PBS). $\ket{D}$ photons pass directly, while $\ket{A}$ photons are routed through a delay line of duration $t_i$ via an optical switch (OS). A half-wave plate (HWP) rotates $\ket{A}$ to $\ket{D}$ before recombination.

2. { $SWAP(P,\mathcal{T}_i)$:} This gate coherently exchanges the polarization bit with the $i$-th time-bin bit, implemented by cascading delay operators and polarization rotations.

3. {$CNOT_{i,P}$:} Realized through the interaction between an NV-center quantum memory and a photon, this gate conditionally flips the photon's polarization based on the memory state~\cite{LRM2017,FB2024,CW2025}.

\section{Quantum-Multiplexed MA QKD Protocols}
\label{sec:mux_QKD}

In this section, we analyze the MA $n$-MUX QKD protocol assuming a single bank of $n+1$ QMs per party. This constraint implies that each transmission must be classically acknowledged before the next begins, providing a clear baseline to illustrate time-bin multiplexing and secret key extraction. We extend this analysis in Section~\ref{sec:QM_banks} to multiple QM banks, enabling pipelined operation and continuous photon emission to maximize throughput in the detector-saturation regime.

The modified Lo–Chau protocol~\cite{LC1999} serves as a foundational entanglement-based QKD scheme, achieving security through entanglement distillation and error detection. However, its implementation is limited by noisy quantum channels used for distributing entangled pairs.

To overcome this challenge, we propose an entanglement-based  MA QKD protocol employing time-bin–multiplexed photons. Building on the entanglement-sharing protocol introduced in the previous subsection, our approach enables the simultaneous establishment of multiple entangled pairs between Alice and Bob while substantially reducing the number of required channel uses.

Specifically, Alice prepares a bank of $n+1$ QMs and applies a sequence of delay operators to generate an $n$-MUX photon that coherently couples to her QMs. This photon is then transmitted to Bob, where it interacts with his local QMs before being measured, creating $n+1$ shared EPR pairs between the two parties. Repeating this process yields the desired number $2N$ of EPR pairs.
Subsequent channel-checking steps, including basis transformations and error estimation, consume $N$ EPR pairs to detect potential eavesdropping. 
Afterwards, the parties apply an $(N,m)$ entanglement distillation protocol~\cite{BDSW96} to generate $m$ nearly perfect EPR pairs from the remaining imperfect ones. 
The distilled pairs are then measured to extract a secure cryptographic key, as formally outlined in Algorithm~\ref{alg:qmuxlochau}.

The primary performance metric in QKD is the \emph{secret key rate}, which quantifies the number of secure bits that can be extracted per channel use.  
Since an $(N,m)$ entanglement distillation protocol is mathematically equivalent to the error correction procedure of a corresponding $[[N,m]]$ Calderbank-Shor-Steane (CSS) code~\cite{CS96,Ste96}, the secret key rate can be analyzed by exploiting the equivalence between entanglement distillation and CSS-code-based error correction, following the Shor–Preskill argument~\cite{SP00}.

\begin{theorem}[Secret Key Rate per Transmitted Photon] \label{thm:lossless_key_rate}
In the lossless setting with perfect sources, detectors, and optical components, 
but  imperfect QMs, the MA $n$-MUX QKD protocol in Algorithm~\ref{alg:qmuxlochau} is secure. 
Its secret key rate per transmitted photon is

\begin{equation}\label{mux_loss_less}
    R_{n} = (n+1) \big[1 - H(E_X) - H(E_Z)\big],
\end{equation}

where $H(\cdot)$ denotes the binary Shannon entropy function, and $E_X$ and $E_Z$ are the intrinsic qubit-flip and phase-flip error rates of the QMs.
\end{theorem}

\begin{proof}[Proof sketch]
We can follow the Shor--Preskill approach~\cite{SP00}, reducing the modified Lo-Chau protocol to a CSS code-based QKD scheme.
We replace the EPR generation with $n$-MUX photons and QMs and assume perfect sources, memories, and detectors.
As in Shor--Preskill~\cite{SP00}, asymptotically the number of secret bits distilled per \emph{qubit} is 
\[
1 - H(E_X) - H(E_Z).
\]
The orthogonality of the multiplexed time-bins and the polarization bit ensures that each carries one independent bit of classical information, so the $n$-MUX photon effectively transmits $n+1$ bits. Hence, the secret key rate per transmitted photon is
$R_n \;=\; (n+1)\big[1 - H(E_X) - H(E_Z)\big].$

\end{proof}

\begin{remark}[Implementation Time]
While Theorem~\ref{thm:lossless_key_rate} characterizes the asymptotic secret key rate per transmitted photon, the achievable wall-clock key rate in practical systems is also constrained by hardware limitations such as detector dead time. For typical single-photon detectors, the recovery time is on the order of $70$--$100$ ns, which limits the maximum effective detection rate in conventional 0-MUX protocols.

In contrast, the $n$-MUX protocol allows multiple entanglement generation attempts to be embedded within a single transmission cycle. Provided that the total duration of the required memory-controlled operations remains comparable to or shorter than the detector recovery time, multiplexing effectively increases the number of usable entanglement attempts per detection window. In practical implementations, the multiplexing depth $n$ can therefore be optimized to balance the increased entanglement yield per photon against the operational overhead.
\end{remark}

\begin{algorithm}[htbp]
\caption{MA $n$-MUX Lo-Chau Protocol}
\label{alg:qmuxlochau}
\begin{algorithmic}[1]
\Require Trusted devices for state preparation and measurement at Alice and Bob.

\State For each single-photon quantum channel between Alice and Bob, Alice prepares $n+1$ QMs in the state $\ket{+}$ and produces an $n$-MUX single photon in $\ket{0}_P \ket{0}_{t_1} \cdots \ket{0}_{t_n}$, while Bob prepares $n+1$ QMs in the state $\ket{+}$.

\State Alice let  the photon sequentially interacts with each QM as in (\ref{eq:CT_interaction}) and then  sends the multiplexed photon to Bob.

\State Bob receives the photon and acknowledges this to Alice. He lets the photon sequentially interact with his $n+1$ local QMs and the degree-of-freedom swap gates as in (\ref{eq:CS_interaction}), then measures the photon and applies the corresponding corrections to the QMs. At this point, Alice and Bob share $n+1$ EPR pairs $\ket{\Phi^{+}}_{AB}$.

\State Repeat the above process until Alice and Bob share $2N$ EPR pairs.

\State  Alice selects a random $2N$-bit string $b$ and applies a Hadamard transform on each QM where $b_i=1$, after receiving acknowledgment from Bob.

\State Alice randomly selects $N$ out of the $2N$ EPR pairs to serve as check bits for detecting Eve's interference.

\State Alice announces the bit string $b$ and which $N$ EPR pairs are the check bits.

\State Bob applies Hadamard gates to the QMs corresponding to the positions  where $b_i=1$.

\State Alice and Bob measure their halves of the $2N$ check EPR pairs in the computational basis and compare results. If too many outcomes disagree, they abort the protocol.

\State Alice and Bob perform an $(N,m)$ entanglement distillation protocol to obtain $m$ nearly perfect EPR pairs.

\State After entanglement distillation, Alice and Bob measure the EPR pairs in the computational basis to obtain the secret key.

\end{algorithmic}
\end{algorithm}

\subsection{Lossy Secret Key Rate Analysis}

In practice, the achievable key rate of the MA $n$-MUX QKD protocol is limited by imperfections in the physical implementation, which critically constrain both the number of multiplexed time bins and the maximum transmission distance.

In the following, we present a theoretical analysis of the secret key rate achievable with our MA $n$-MUX QKD protocol, taking into account realistic error sources such as memory imperfections, channel loss, and detector dark counts. 
This analysis further allows us to determine the maximum feasible multiplexing depth at long distances, such as $L=300$ km, under experimentally realistic conditions.

For illustration, we assume that the system bandwidth allows only a single photon to be transmitted at a time; the analysis can be directly generalized to larger bandwidths.

The probability that a single-photon signal is successfully transmitted and detected is~\cite{NMW2017}

\begin{equation}\label{eq:Pho_arrive_prob}
    P_{s} = \eta_{d}\eta_{s}\eta_{os}^{3n} e^{-L/L_{att}}, 
\end{equation}
where $\eta_s$, $\eta_d$, and $\eta_{os}$ are the efficiencies of the single-photon source, the detector, and each optical switch, respectively. Here, we assume all the $CNOT_{i,P}$ operation are perfect. The exponential term $e^{-L/L_{att}}$ accounts for channel loss, with $L$ the fiber distance and $L_{att}$ the attenuation length.

We assume that photon loss in short optical delay lines and imperfections in the photon–microcavity interaction are negligible compared to fiber loss over long distances. 
However, imperfections in the QM–photon interaction can lead to errors. In particular, if the controlled interaction fails, the photon may propagate without undergoing the intended conditional operation, resulting in an incorrect correlation between the photon and the quantum memory. This contributes to the quantum bit error rate (QBER).
In our analysis, such imperfections are modeled as depolarizing noise. Within the Shor–Preskill framework, these errors are accounted for through the entanglement distillation procedure.

In our construction, each delay operator contains one optical switch, and each degree-of-freedom swap gate contains two. Therefore, to establish $n+1$ EPR pairs using an $n$-MUX photon, the photon traverses a total of $3n$ optical switches. This results in the factor $\eta_{os}^{3n}$ in the transmission probability $P_s$ in~\eqref{eq:Pho_arrive_prob}.

Next, we would like to recall that in the practical setting, the intrinsic error rates $E_X$ and $E_Z$ are determined solely by the imperfections of the QMs:
\begin{align}
    E_X =& \tfrac{1}{2}\left(1 - e^{-\tau/T_1}\right), \label{eq:EXp}\\
    E_Z =& \tfrac{1}{2}\left(1 - e^{-\tau/T_2}\right),\label{eq:EZp}
\end{align}
where $T_1$ and $T_2$ denote the relaxation and coherence times of the QMs, respectively, and $\tau$ is the storage duration of each EPR in the QMs. Longer storage durations will lead to higher error rates.

Detector dark counts are spurious detection events that occur in the absence of a photon, while a detector click refers to any registered detection event, including both true photon arrivals and dark counts. 
Since dark counts do not generate entanglement between Alice and Bob, their effect is modeled as a random guess with a $50\%$ error probability. 
To account for dark counts, the intrinsic error rates $E_X$ and $E_Z$ should be modified as~\cite{MQZL2005}:
\begin{equation}\label{dc_ER}
    E'_{\alpha} = \frac{E_{\alpha} P_{s} + \tfrac{1}{2} (1-P_{s}) P_{dc}}{P_{click}}, 
    \quad \alpha \in \{X,Z\},
\end{equation}
where $P_{dc}$ is the probability of a dark count, and $P_{click} = P_{s} + (1-P_{s}) P_{dc}$ is the total probability of a detector click. 

With these definitions, we can now state the main result as follows.

\begin{theorem}[Lossy Secret Key Rate]\label{thm:final_keyrate}
The secret key rate per transmitted photon of the MA $n$-MUX QKD protocol in the presence of channel loss and imperfect devices is
\begin{equation}\label{mux_final_key_theorem}
    R_{n} = (n+1) P_{click}\big[1 - H(E'_X) - H(E'_Z)\big],
\end{equation}
where $E_X'$ and $E_Z'$ are the bit-flip and phase-flip error rates corrected for dark counts as defined in~\eqref{dc_ER}.
\end{theorem}

\begin{proof}
Each multiplexed photon encodes $n+1$ independent bits, giving a raw key rate of 
\[
R' = (n+1) \big[1 - H(E'_X) - H(E'_Z)\big].
\]  
Equation~\eqref{eq:Pho_arrive_prob} defines the probability $P_s$ of a photon successfully passing through the system, and including dark counts gives the overall click probability $P_{click}$. 
Multiplying the raw key rate $R'$ by $P_{click}$ yields the secret key rate stated in \eqref{mux_final_key_theorem}.
\end{proof}

\begin{table}[ht]
    \centering
    \caption{System parameters and loss modeling assumptions.}
    \label{tb:parameter}
    \renewcommand{\arraystretch}{1.2}
    \setlength{\tabcolsep}{8pt}
    {\footnotesize
        \begin{tabular}{p{5.2cm} p{2cm}}
            \toprule
            \textbf{Component / Parameter} & \textbf{Value / Model} \\
            \midrule
            \multicolumn{2}{l}{\textit{Lossy Components}} \\
            Fiber attenuation & 0.2 dB/km \\
            \quad Attenuation length, $L_{att}$ & 22 km \\
            Photon source efficiency, $\eta_s$ & 71.2\%~\cite{Ding2025} \\
            Detector efficiency, $\eta_d$ & 98\%~\cite{Hu:20,Reddy:20} \\
            \quad Dark count probability, $P_{dc}$ & $10^{-8}$ (1 Hz~\cite{Reddy:20}) \\
            Optical switch efficiency, $\eta_{os}$ & 99\%~\cite{LMN19} \\
            \midrule
            \multicolumn{2}{l}{\textit{Quantum Memory Decoherence}} \\
            Relaxation time, $T_1$ & 10 s~\cite{BG2013} \\
            Coherence time, $T_2$ & 0.5 s~\cite{BG2013} \\
            \midrule
            \multicolumn{2}{l}{\textit{Ideal Components (Negligible Loss)}} \\
            QM-photon $CNOT_{i,P}$ operations & Ideal \\
            Local delay lines $\mathcal{T}_i$ & Ideal \\
            Degree-of-freedom swap $SWAP(P,\mathcal{T}_i)$ & Ideal \\
            \midrule
            \multicolumn{2}{l}{\textit{Other Parameters}} \\
            Speed of light in fiber, $c$ & $2 \times 10^{8}$ m/s \\
            Source repetition rate & 50 MHz~\cite{RV2011} \\
            \bottomrule
        \end{tabular}
    }
\end{table}

 In practical implementations, the protocol is subject to timing jitters and fiber dispersion. Detector jitter may lead to the misidentification of time-bin indices, while dispersion causes wavepacket broadening and overlap between adjacent bins. In our model, these imperfections are conservatively treated as sources of bit-flip and phase errors. These errors are effectively mitigated by the entanglement distillation process, which purifies the initial state fidelity at the cost of a reduced raw entanglement generation rate.

\subsection{Sufficient bandwidth for multiplexing}

To derive an upper bound on the optimal multiplexing depth $n$, we consider an idealized scenario in which a sufficiently large number of photon transmission channels (approximately $2N/n$) are available. This allows all required EPR pairs to be generated simultaneously, so that each pair experiences the same storage time in the quantum memories and hence the same noise level. 
This assumption simplifies the analysis by avoiding variations in decoherence across different EPR pairs.

We define the secret key rate ratio $r_{n} = R_{n}/R_{n-1}$ and use it to estimate the maximum feasible multiplexing depth. 
Specifically, we define the maximum multiplexing depth as the largest $n$ such that $r_{n} > 1$ at $L = 300\,\text{km}$, which we take as a representative distance for current QKD implementations.

\begin{proposition} \label{prop:mux_depth}
In a lossy MA $n$-MUX QKD protocol with sufficient bandwidth, 
the maximum achievable multiplexing depth at a transmission distance of $L=300$~km is $n=27$, 
given the experimental parameters summarized in Table~\ref{tb:parameter}.
\end{proposition}

\begin{figure}[h]
  \centering
  \includegraphics[width=0.85\linewidth]{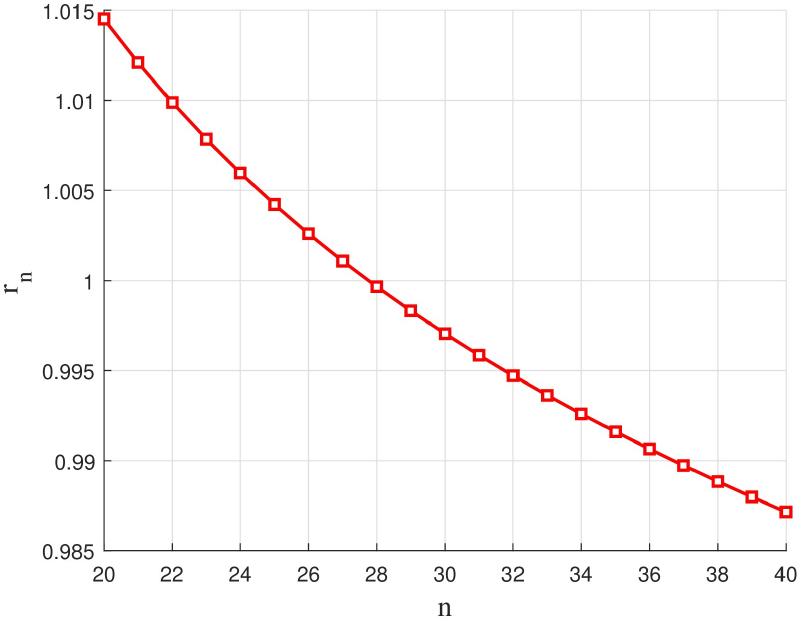}
\caption{Secret key rate ratio $r_n$ as a function of the multiplexing depth $n$ at $L=300$ km. The ratio falls below unity for $n \geq 28$.}
  \label{fig:KeyRateRatio}
\end{figure}

\begin{proof}
	Before the entanglement distillation in Step~10 of Algorithm~\ref{alg:qmuxlochau}, the total communication time cost is $3L/c$: one round for Alice to transmit a multiplexed photon to Bob, one round for Bob to return an acknowledgment, and one round for Alice to announce the string $b$ along with the positions of the check bits. If Bob reports that no photon was received, the protocol is restarted.  
Hence, we set $\tau = 3L/c$ when estimating $E_X$ and $E_Z$.  

Figure~\ref{fig:KeyRateRatio} shows the ratio $r_n$ as a function of the multiplexing depth $n$. The ratio decreases monotonically and falls below one for $n \geq 28$. Therefore, the maximum feasible multiplexing depth is $n=27$.  
\end{proof}

For comparison, we analyze the secret key rate of a conventional MA QKD protocol without multiplexed photons (i.e., 0-MUX). 
In this case, no optical switches are involved, so $\eta_{\mathrm{os}}=1$. 
The photon transmission and detection probabilities simplify to
\begin{align}
    P'_{s} &= \eta_{d}\eta_{s} e^{-L/L_{\mathrm{att}}}, \label{eq:Conv_Pho_arrive_prob} \\   
    P'_{\mathrm{click}} &= P'_{s} + \big(1-P'_{s}\big) P_{dc}. \label{eq:Conv_click_prob}
\end{align}

To ensure a fair comparison, we assume the same overall transmission bandwidth, so that 2N/(n+1) photon transmission channels are available. Compared with the simultaneous generation enabled by the $n$-MUX protocol, this setting significantly increases the waiting time of the EPR pairs stored in the QMs. Earlier pairs must be stored until the entire batch is collected, which consequently increases the effective bit-flip and phase-flip error rates in the QMs.

For analytical simplicity, we assume that each photon experiences an expected waiting time. Accordingly, the $2N$ transmitted photons can be partitioned into $n+1$ groups, which in turn generate $n+1$ sets of EPR pairs. All EPR pairs within the same group undergo identical error rates in their corresponding QMs and therefore contribute equally to the secret key rate.

The lossless key rate for the $i$-th group of EPR pairs, as given by Theorem~\ref{thm:lossless_key_rate}, is
\begin{equation}
    R^{(i)} = 1 - H\!\left(E^{(i)}_{X}\right) - H\!\left(E^{(i)}_{Z}\right),
\end{equation}
where $E^{(i)}_{X}$ and $E^{(i)}_{Z}$ are the bit-flip and phase-flip error rates corresponding to the EPR pairs in group $i$.

Accounting for dark counts, the effective error rates for the $i$-th group are
\begin{equation}
    E'^{(i)}_{\alpha} 
    = 
    \frac{ E^{(i)}_{\alpha} P'_{s} + \tfrac{1}{2}\big(1-P'_{s}\big)P_{dc} }{ P'_{click} }, 
    \quad \alpha \in \{X,Z\}.
\end{equation}
Hence, the secret key contribution per transmission for the $i$-th group is
\begin{equation}
    R'^{(i)} = P'_{click}\Big[1 - H\!\big(E'^{(i)}_{X}\big) - H\!\big(E'^{(i)}_{Z}\big)\Big].\label{eq:Rprime_i}
\end{equation}

Next, we estimate the storage duration of an EPR pair in each transmission group. 
If a photon transmission fails, the corresponding EPR attempt restarts; thus, the expected time to establish one EPR pair is $\tfrac{2L}{c}$, corresponding to Alice sending a photon and waiting for Bob’s confirmation of successful arrival. 
In addition, an EPR pair in the $i$-th group must wait for the subsequent $(n+1-i)$ groups to complete, which requires on average $(n+1-i)\tfrac{2L}{cP'_s}$. 
Finally, all EPR pairs incur an additional communication delay of $\tfrac{L}{c}$ due to Alice’s classical announcements. 
Combining these contributions, the total storage duration of an EPR pair in group $i$ is
\begin{equation}\label{eq:time_cost}
    \tau_i = \frac{2L}{c} + (n+1-i)\frac{2L}{cP'_s}+\frac{L}{c} ,
\end{equation}
for $i = 1, \dots, n+1$.

\begin{theorem}[Conventional Secret Key Rate]\label{thm:final_keyrate_Conv}
For the conventional (0-MUX) MA Lo–Chau QKD protocol, the secret key rate per channel use over $n+1$ sequential transmissions is
\begin{equation}\label{eq:noQMX_KR_correct}
    R^{C}_{n+1} = \frac{1}{n+1}\sum_{i=1}^{n+1} R'^{(i)}.
\end{equation}
\end{theorem}

\begin{proof}
For each channel use, the expected key contribution is $R'^{(i)}$ in~\eqref{eq:Rprime_i}, which already accounts for transmission probability, dark counts, and memory-induced errors. The total expected key bits over $n+1$ uses is then $\sum_{i=1}^{n+1} R'^{(i)}$. Normalizing it by $n+1$ yields the per-use rate in~\eqref{eq:noQMX_KR_correct}.
\end{proof}

\begin{figure}[htbp]
    \centering
    \includegraphics[width=0.9\linewidth]{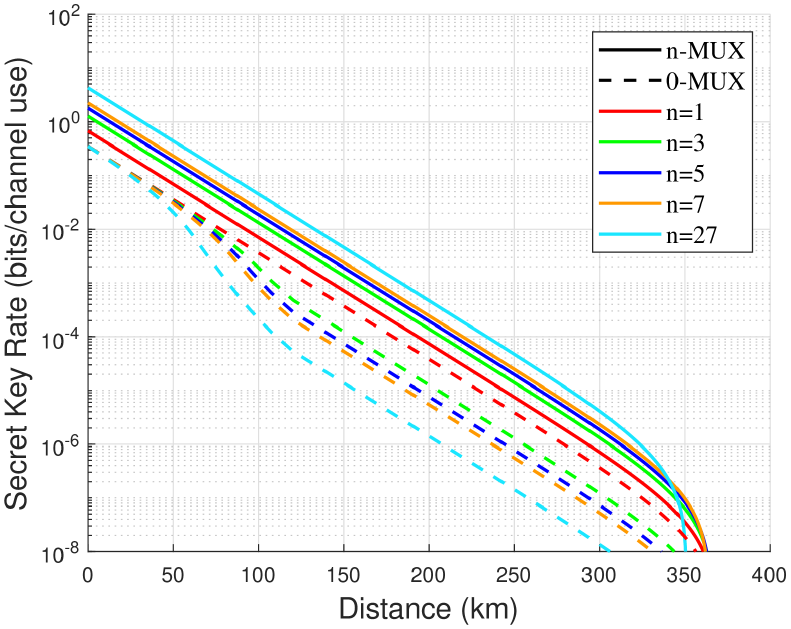}
    \caption{Secret key rates for the MA $n$-MUX QKD protocol compared with the conventional 0-MUX protocol, assuming a bandwidth of $2N/(n+1)$. Dotted lines show the performance of the conventional (0-MUX) protocol under the same bandwidth constraint as the $n$-MUX protocol.}
    \label{fig:lochau_keyrate_log}
\end{figure}

We numerically evaluate the key rate under the parameters specified in Table~\ref{tb:parameter}. To highlight the benefit of multiplexing, we compare the performance of our MA $n$-MUX QKD protocol with the conventional 0-MUX protocol. The secret key rates predicted by Theorems~\ref{thm:final_keyrate} and~\ref{thm:final_keyrate_Conv} are plotted in Figure~\ref{fig:lochau_keyrate_log} as functions of the transmission distance $L$ for various multiplexing depths $n$.

The results in Figure~\ref{fig:lochau_keyrate_log} show that using $n$-MUX photons yields an approximately linear increase in the secret key rate as $n$ grows, while the 0-MUX protocol suffers from reduced performance due to the decreasing available bandwidth $2N/n$. For the 0-MUX protocol, the curves corresponding to different $n$ nearly coincide for transmission distances below $60$ km.

Furthermore, for a fixed multiplexing depth $n$, our protocol achieves substantially longer transmission distances at the same key rate. For instance, at a rate of $10^{-4}$ with $n=7$, the $n$-MUX protocol supports $217.8$ km, compared to less than $133$ km for 0-MUX.

These results highlight that entanglement-based QKD protocols relying on entanglement distillation are strongly limited by the bandwidth of photon transmission channels, and multiplexing offers a practical route to overcoming this constraint.

\subsection{Limited Bandwidth for Multiplexing}

We now consider a more restricted scenario in which the number of photons that can be transmitted simultaneously is limited to a small constant, representing a practical bandwidth constraint.

To illustrate the advantage of multiplexing under this bandwidth constraint, we assume 10 photons can be sent simultaneously and compare the performance of our MA $n$-MUX QKD protocol with the conventional 0-MUX scheme.

Suppose Alice and Bob aim to establish a total of $N = 1400$ EPR pairs for entanglement distillation. In the conventional 0-MUX scheme, each photon generates only one EPR pair, so the bandwidth limitation allows at most 10 EPR pairs per transmission round. In contrast, the $n$-MUX QKD protocol with multiplexing depth $n = 27$ (theoretical maximum from Proposition~\ref{prop:mux_depth}) allows each multiplexed photon to generate 28 EPR pairs. Consequently, with a bandwidth of 10 photons per round, up to 280 EPR pairs can be created per round, providing a substantial increase in effective throughput.

To quantitatively compare the performance, we evaluate the average secret key rates of the two protocols.

For the $n$-MUX protocol, the $N$ photons are divided into $\frac{N}{10(n+1)}$ groups. The key rate contributed by group $i$ is
\begin{equation}
R_{n}^{(i)} = (n+1) P_{click} \big[ 1 - H(E'^{(i)}_{X}) - H(E'^{(i)}_{Z}) \big], 
\end{equation}
for $i = 1, \dots, \frac{N}{10(n+1)},$
where the error rates $E'^{(i)}_{X}$ and $E'^{(i)}_{Z}$ are adjusted according to the storage duration, which can be derived similarly to~\eqref{eq:time_cost} by replacing $n$ with $\frac{N}{10(n+1)}$.

The resulting lossy secret key rate per photon transmission for the $n$-MUX protocol is
\begin{equation}
R^{10}_{n} = \frac{10(n+1)}{N} \sum_{i=1}^{\frac{N}{10(n+1)}} R_{n}^{(i)},
\end{equation}
where the superscript $10$ denotes the bandwidth.

Similarly, for the conventional 0-MUX protocol, the key rate per photon transmission is
\begin{equation}
R^{C,10}_{n} = \frac{10}{N} \sum_{i=1}^{N/10} R^{(i)}.
\end{equation}

\begin{figure}[htbp]
    \centering
    \includegraphics[width=0.9\linewidth]{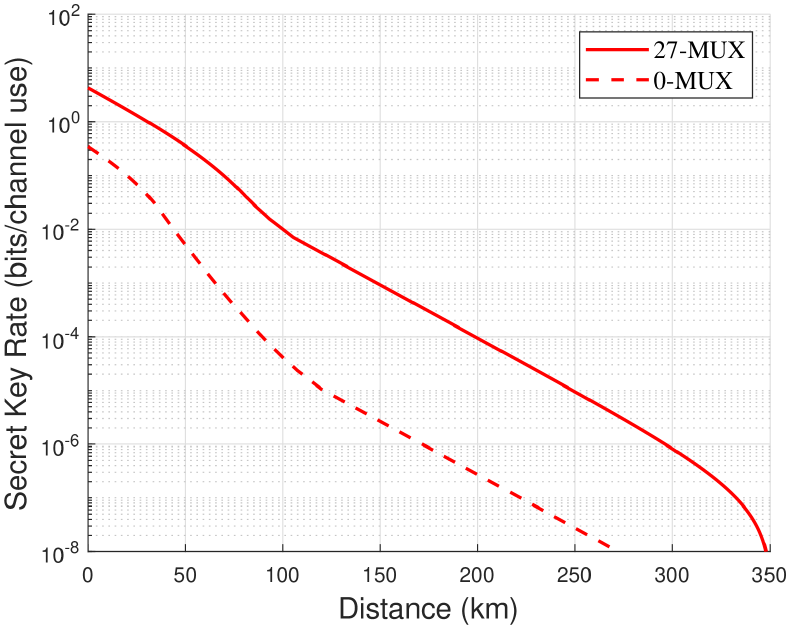}
    \caption{Comparison of secret key rates for the conventional 0‑MUX and 27‑MUX MA QKD protocols under a bandwidth limit of 10 photon channels. The 27‑MUX protocol achieves higher key rates and longer transmission distances.}
    \label{fig:bandwidth}
\end{figure}

Figure~\ref{fig:bandwidth} compares the conventional 0‑MUX and 27‑MUX protocols. The red solid line represents the 27‑MUX protocol, and the red dashed line represents the 0‑MUX protocol, showing the higher key rate and longer transmission distance achievable with the $n$‑MUX protocol.

\section{Time-Bin Multiplexed measurement-device-independent QKD}
\label{sec:mux_MDIQKD}

In this section, we present an entanglement-based MDI-QKD protocol 
employing time-bin multiplexing with an untrusted intermediate node. 

We first introduce the protocol using a single memory bank to establish 
the security framework (Section~\ref{sec:MDI_single}), then extend 
to the memory-insensitive (MI) architecture with multiple banks 
(Section~\ref{sec:QM_banks}), which represents the practical 
implementation suitable for realistic quantum networks.

\subsection{MDI-QKD with a Single QM Bank }
\label{sec:MDI_single}

In the MDI-QKD setting, the two communicating parties, Alice and Bob, each possess trusted devices capable of preparing entangled states between local qubits and a single time-bin multiplexed photon. The photon carries no information prior to leaving the trusted laboratories. The intermediate node, Charlie, is entirely untrusted and may perform arbitrary quantum operations and classical post-processing. No assumptions are made regarding his measurement devices, which may even be under adversarial control. 
Extending Algorithm~\ref{alg:qmuxlochau} to this setting, we obtain our entanglement-based MA $n$-MUX MDI-QKD protocol, described in Algorithm~\ref{alg:mp-sdiqkd_repeater}.

\begin{algorithm}[htbp]
\caption{ MA $n$-MUX MDI-QKD Protocol}
\label{alg:mp-sdiqkd_repeater}
\begin{algorithmic}[1]
\Require Trusted devices for state preparation and measurement at Alice and Bob; QMs available at all nodes.

\State \textbf{Alice--Charlie entanglement:} 
For each single-photon quantum channel between Alice and Charlie, Alice prepares QM $1,3,\dots,2n+1$ in $\ket{+}$ and generates an $n$-MUX single photon in $\ket{0}_P\ket{0}_{t_1}\dots\ket{0}_{t_n}$, while Charlie prepares QM $1',3',\dots,(2n+1)'$ in $\ket{+}$.

\State  
Alice lets the photon sequentially interact with each QM as in (\ref{eq:CT_interaction}) and sends the multiplexed photon to Charlie.

\State  
Upon receiving the photon, Charlie acknowledges the reception to Alice and Bob, lets the photon sequentially interact with his $n+1$ QMs and the degree-of-freedom swap gates as in (\ref{eq:CS_interaction}), then measures the photon and applies the corresponding corrections. Alice and Charlie then share $n+1$ EPR pairs $\ket{\Phi^{+}}_{AC}$.

\State \textbf{Bob--Charlie entanglement:} 
Similarly, Bob prepares QM $2,4,\dots,2n+2$ in $\ket{+}$ and emits a photon that sequentially interacts with his QMs before sending the multiplexed photon to Charlie until success is achieved. 
Charlie prepares QM $2',4',\dots,(2n+2)'$ in $\ket{+}$ and lets the photon sequentially interact with these $n+1$ QMs and the degree-of-freedom swap gates, then measures the photon and applies the corresponding corrections. Bob and Charlie then share $n+1$ EPR pairs $\ket{\Phi^{+}}_{BC}$.

\State \textbf{Entanglement swapping:} 
Once Charlie holds one entangled QM with Alice and one with Bob, he performs a Bell measurement on QMs $(2j+1)'$ and $(2j+2)'$, creating entanglement between Alice's QM $2j+1$ and Bob's QM $2j+2$.

\State \textbf{Pauli correction:} 
Bob applies the corresponding corrections according to the measurement outcomes.

\State Repeat the above process until Alice and Bob share $2N$ EPR pairs.

\State \textbf{Channel check:} 
Alice selects a random $2N$-bit string $b$ containing $N$ ones and $N$ zeros, then announces the pairs used for channel checks. For each qubit $i$ where $b_i=1$, Alice and Bob apply a Hadamard before measurement.

\State  
Alice and Bob measure their halves of the $2N$ check EPR pairs in the computational basis and compare results. If too many outcomes disagree, they abort the protocol.

\State Alice and Bob perform an $(N,m)$ entanglement distillation protocol to obtain $m$ nearly perfect EPR pairs.

\State After distillation, Alice and Bob measure the EPR pairs in the computational basis to obtain the secret key.

\end{algorithmic}
\end{algorithm}

\begin{proposition}[Security of MA $n$-MUX MDI-QKD]
The protocol in Algorithm~\ref{alg:mp-sdiqkd_repeater} is secure in the composable framework.
\end{proposition}

\begin{proof}[Justification]
The security follows from the indistinguishability of the entanglement-based MDI-QKD protocol and the original MDI-QKD~\cite{LCQ12,GLLP04} from the adversary’s perspective.  
In particular, Eve cannot distinguish the two protocols in the output key distribution~\cite{BBM92,GLLP04}.  
Since composable security of MDI-QKD is established, the same guarantee extends to our MA $n$-MUX MDI-QKD protocol.  
\end{proof}

Now, we analyze the secret key rate of the MDI-QKD protocol assuming a transmission bandwidth of 2N/(n+1) photons. We consider Alice and Bob separated by a distance $L$, with Charlie positioned midway at a distance $L/2$ from each. The setup resembles the multiplexed Lo-Chau protocol; however, here Charlie, acting as the central node, sequentially establishes entangled links with Alice and Bob. 

First, Alice must attempt to establish $n+1$ EPR pairs $\ket{\Phi^+}$ with Charlie repeatedly until successful entanglements are achieved. After receiving Charlie's acknowledgment, Bob proceeds to generate additional $n+1$ EPR pairs in the state $\ket{\Phi^{+}}$ shared between Charlie and himself. Alice will announce the bit string $b$ and the positions of check bits after the swapping. Thus, the time cost associated with storing the entanglement between Alice and Charlie ($t^{A}$), as well as Bob and Charlie ($t^{B}$), is the following:
\begin{equation}\label{eq:MDI_time_cost}
    \begin{aligned}
        &t^{A} = 2 \frac{L/2}{c} + \frac{L}{cP_{s}}+\frac{L}{c},\\
        &t^{B} = 2\frac{L/2}{c}+\frac{L}{c}. 
    \end{aligned}
\end{equation}
Now, we define the error rate of the EPR pairs between Alice and Charlie as follows:
\begin{equation}
    \begin{aligned}
        E^{AC}_X =& \tfrac{1}{2}\left(1 - e^{-t^{A}/T_1}\right),\\
        E^{AC}_Z =& \tfrac{1}{2}\left(1 - e^{-t^{A}/T_2}\right),
    \end{aligned}
\end{equation}
then, the error rate of the EPR pairs between Bob and Charlie is the following:
\begin{equation}
    \begin{aligned}
        E^{BC}_X =& \tfrac{1}{2}\left(1 - e^{-t^{B}/T_1}\right),\\
        E^{BC}_Z =& \tfrac{1}{2}\left(1 - e^{-t^{B}/T_2}\right),
    \end{aligned}
\end{equation}
here, utilize~\eqref{dc_ER} to correct $E^{AC}_{\alpha}$ and $E^{BC}_{\alpha}$ ($\alpha \in \{X,Z\}$) to get $E'^{AC}_{\alpha}$ and $E'^{BC}_{\alpha}$. We need to merge $E'^{AC}_{\alpha}$ and $E'^{BC}_{\alpha}$ to $E'^{AB}_{\alpha}$ for entanglement swapping. 

\begin{theorem}[Error composition under entanglement swapping]\label{thm:final_keyrate_MDI}
Consider two elementary links, Alice--Charlie (A--C) and Bob--Charlie (B--C), each described by independent Pauli channels with bit-flip error rates $E_X'^{AC}, E_X'^{BC}$ and phase-flip error rates $E_Z'^{AC}, E_Z'^{BC}$, respectively. After a Bell measurement at the relay node Charlie, the effective errors on the entangled state shared between Alice and Bob are given by
\begin{align}
E_X'^{AB} &= E_X'^{AC} + E_X'^{BC} - 2 E_X'^{AC} E_X'^{BC}, \label{eq:ex_ab}\\
E_Z'^{AB} &= E_Z'^{AC} + E_Z'^{BC} - 2 E_Z'^{AC} E_Z'^{BC}. \label{eq:ez_ab}
\end{align}
\end{theorem}

\begin{proof}
Take the bit-flip errors as an example. In the entanglement swapping procedure, the Bell measurement projects the two links A--C and B--C into an effective A--B entangled state. A bit-flip on either link leads to an effective bit-flip on A--B, while simultaneous bit-flips on both links cancel each other (since $X \otimes X$ acts trivially on the Bell basis). Therefore, the total bit-flip error rate is
\[
E_X'^{AB} = E_X'^{AC}(1-E_X'^{BC}) + (1-E_X'^{AC}) E_X'^{BC},
\]
which simplifies to~\eqref{eq:ex_ab}. The same reasoning applies to phase-flip errors, yielding~\eqref{eq:ez_ab}.
\end{proof}

Now, we have the secret key rate as same as Theorem~\ref{thm:final_keyrate}
\begin{equation}\label{eq:MDI_KeyRate}
    R_{S} = (n+1) P_{click}\big[1 - H(E'^{AB}_X) - H(E'^{AB}_Z)\big].
\end{equation}

 \subsection{Pipelined Operation with Multiple Memory Banks}
\label{sec:QM_banks}

We now consider the case where multiple QM banks are available at each node, following the memory-insensitive (MI) framework~\cite{CK2007}. 
To overcome the long waiting times inherent in the single-bank architecture, we employ a pipelined scheme in which multiple independent QM banks enable simultaneous entanglement generation attempts.
This allows the system to process many incoming photons in a pipelined fashion, avoiding the long waiting times associated with single-bank protocols. 

Each memory bank consists of $n+1$ quantum memories that operate independently and can be used to generate one set of $n+1$ EPR pairs.
A high-repetition-rate single-photon source emits a pulse train of $k$ photons. 
By synchronizing the emission time with the memory addressing, the $i$-th photon is directed to the $i$-th QM bank ($i=1,\dots,k$). 

Each photon is therefore processed by a different QM bank, allowing the system to perform $k$ independent entanglement-generation attempts within one pulse train without increasing the storage time of previously generated EPR pairs. The detailed procedure is summarized in Algorithm~\ref{alg:mI-sdiqkd_repeater}.

\begin{algorithm}[htbp]
\caption{MI $n$-MUX MDI-QKD Protocol}
\label{alg:mI-sdiqkd_repeater}
\begin{algorithmic}[1]

\State \textbf{Alice--Charlie entanglement:} 
For each quantum channel, Alice and Charlie prepare $k$ QM banks, each containing $n+1$ QMs initialized in $\ket{+}$. Alice generates a pulse train of $k$ $n$-MUX photons in $\ket{0}_P\ket{0}_{t_1}\cdots\ket{0}_{t_n}$.

\State 
Following~(\ref{eq:CT_interaction}), Alice sequentially interacts the $i$-th photon with the $i$-th QM bank and sends it to Charlie.

\State 
Charlie acknowledges the reception. Following~(\ref{eq:CS_interaction}), he interacts the $i$-th photon with the $i$-th QM bank, performs the swap operations, then measures the photon and applies the corresponding QM corrections, establishing $n+1$ EPR pairs $\ket{\Phi^{+}}_{AC}$.

\State \textbf{Bob--Charlie entanglement:} 
Similarly, Bob prepares $k$ QM banks (each with $n+1$ QMs) and a pulse train of $k$ $n$-MUX photons. After interacting the $i$-th photon with the $i$-th QM bank, he sends it to Charlie, who repeats the interaction, measurement, and correction procedure to establish $n+1$ EPR pairs $\ket{\Phi^{+}}_{BC}$.

\State \textbf{Entanglement swapping:} 
Once Charlie shares at least one entangled QM with both Alice and Bob, he performs a Bell measurement on the corresponding QM pair. Remaining entangled QMs from the same transmission window are discarded, and the outcomes are broadcast.

\State \textbf{Pauli correction:} 
Bob applies the corresponding QM corrections according to Charlie's broadcast outcomes.

\State Repeat the above process until Alice and Bob share $2N$ EPR pairs.

\State \textbf{Channel check:} 
Alice selects a random $2N$-bit string $b$ containing $N$ ones and $N$ zeros, then announces the pairs used for channel checks. For each qubit $i$ where $b_i=1$, Alice and Bob apply a Hadamard before measurement.

\State 
Alice and Bob measure their halves of the $2N$ check EPR pairs in the computational basis and compare results. If too many outcomes disagree, they abort the protocol.

\State Alice and Bob perform an $(N,m)$ entanglement distillation protocol to obtain $m$ nearly perfect EPR pairs.

\State After distillation, Alice and Bob measure the EPR pairs in the computational basis to obtain the secret key.

\end{algorithmic}
\end{algorithm}

For the correct operation of the protocol, it suffices that at least one of the $k$ attempts successfully establishes an entangled link. Consequently, we do not utilize the residual entanglement discussed in~\cite{CK2007}; any additional successful connections generated within the same transmission window are considered redundant and are discarded in our rate analysis.

Compared with the single-attempt protocol, the memory-insensitive approach utilizes the pulse train to significantly boost the success probability per transmission window. 
The modified key rates are obtained by replacing $P_s$ with $P_s^{(k)}$ in Theorem~\ref{thm:final_keyrate_MDI}.
Specifically, the probability that at least one photon in a train of $k$ pulses is successfully transmitted and detected is  
\begin{equation}\label{eq:boosted_Ps} 
    P^{(k)}_{s} = 1 - (1 - P_{s})^k, 
\end{equation}
where $P_{s}$ is defined in~\eqref{eq:Pho_arrive_prob}.
Similarly, the probability $P_s$ appearing in~\eqref{eq:MDI_time_cost} is replaced by $P_s^{(k)}$.
The corresponding error rates $E_X'^{AB}$ and $E_Z'^{AB}$ are denoted by $\mathcal{E}_X'^{AB}$ and $\mathcal{E}_Z'^{AB}$, respectively.

In the memory-insensitive architecture, the dark-count probability must also be adjusted because the detector is exposed to $k$ temporal modes. 
The effective dark-count probability $P_{dc}^{(k)}$, representing the probability that at least one dark count occurs across the $k$ pulses, is
\begin{equation}\label{eq:boosted_dark_count_P_dc}
    P_{dc}^{(k)} = 1 - (1 - P_{dc})^{k}.
\end{equation}

Accordingly, $P_s$ in the click probability is replaced by~\eqref{eq:boosted_Ps}, and $P_{dc}$ is replaced by~\eqref{eq:boosted_dark_count_P_dc}. The resulting click probability is denoted by $P_{click}^{(k)}$.

\begin{theorem}[Secret Key Rate of Memory-Insensitive Protocol]
For the memory-insensitive protocol, the secret key rate per channel use over $k$ sequential transmissions is
\begin{equation}
    R^{(k)}_S = (n+1) P_{click}^{(k)} \left[1-H(\mathcal{E}'^{AB}_X)-H(\mathcal{E}'^{AB}_Z)\right].
\end{equation}
\end{theorem}

\begin{proof}
Alice and Bob each transmit a pulse train of length $k$ per communication round. Under time-bin multiplexing, a single successful heralding event within this window establishes $n+1$ EPR pairs.
 The probability that at least one detector click occurs  across the $k$ pulses either from a true detection or a dark count  is
\begin{equation*}
    P_{click}^{(k)} = P_s^{(k)} + \bigl(1 - P_s^{(k)}\bigr) P_{dc}^{(k)} .
\end{equation*}

The resulting $n+1$ EPR pairs are subject to effective bit-flip and phase-flip errors, denoted by $\mathcal{E}_X'^{AB}$ and $\mathcal{E}_Z'^{AB}$. These parameters incorporate both the reduced memory storage time due to the boosted success probability $P_s^{(k)}$ and the increased noise floor caused by the effective dark-count probability $P_{dc}^{(k)}$.

Following Theorem~\ref{thm:lossless_key_rate}, the number of secure bits distilled per EPR pair is $1-H(\mathcal{E}_X'^{AB})-H(\mathcal{E}_Z'^{AB})$. Hence the secret key rate is
\begin{equation*}
    R_S^{(k)} = (n+1) P_{click}^{(k)} \left[1-H(\mathcal{E}_X'^{AB})-H(\mathcal{E}_Z'^{AB})\right].
\end{equation*}
\end{proof}

\begin{figure}[htbp] 
    \centering 
    \begin{subfigure}[b]{0.4\textwidth}
        \centering
        \includegraphics[width=\linewidth]{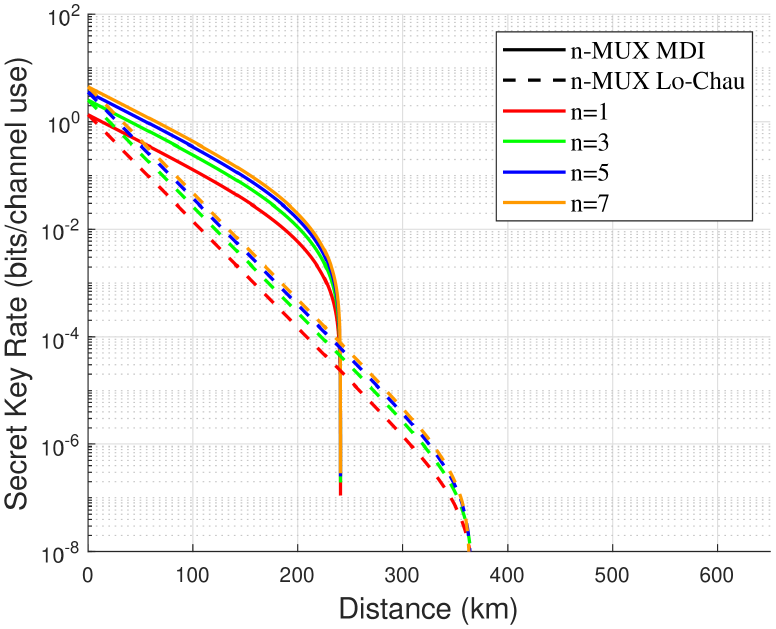}
        \caption{Secret key rate without MI protocol ($k$=1)}
        \label{fig:MDI_VS_noMDI}
    \end{subfigure}
    \hfill 
    \begin{subfigure}[b]{0.4\textwidth}
        \centering
        \includegraphics[width=\linewidth]{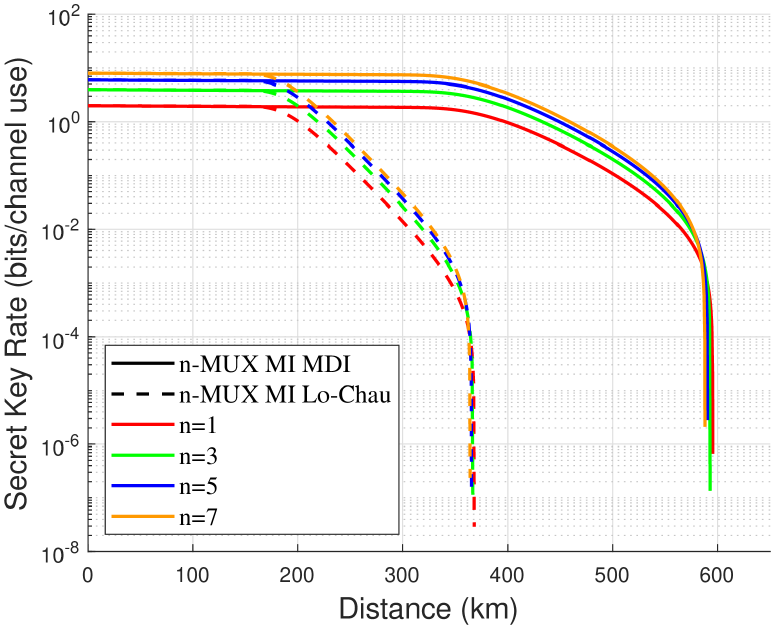}
        \caption{Secret key rate with MI protocol ($k$=10000)}
        \label{fig:MDI_VS_noMDI_MI}
    \end{subfigure}
    \caption{Performance of the MI $n$-MUX MDI-QKD protocol assuming 2N/(n+1) photon transmission channels. (a) Secret key rate without MI protocol ($k$=1). (b) Secret key rate with MI protocol ($k$=10000). In both subfigures, solid lines show the results with an untrusted quantum repeater, while dashed lines correspond to the case without a repeater.
    The 
effective transmission distance (where rate drops to 90\% of peak) 
nearly doubles from 169–173 km to 324–331 km with the repeater.
    }
\end{figure}

As illustrated in Figure~\ref{fig:MDI_VS_noMDI}, we compare the secret
key rates with and without an untrusted quantum repeater using the
single-bank protocol ($k=1$).

Equation~\eqref{mux_final_key_theorem} (dashed line) represents direct
transmission, while Equation~\eqref{eq:MDI_KeyRate} (solid line) shows
the performance with a repeater. The repeater-assisted MDI protocol
achieves a higher key rate for distances below approximately 240 km.
However, the key rate rapidly vanishes beyond this point and does not
exhibit the doubled transmission distance typically expected from a
repeater. This behavior arises because the single-bank protocol suffers
from long waiting times: entanglement must be sequentially established
on the A–C and B–C links, increasing the effective storage time and
therefore the accumulated errors due to the finite memory coherence
time ($T_2$). This coherence-limited regime motivates the use of the
memory-insensitive architecture with multiple banks.

As shown in Figure~\ref{fig:MDI_VS_noMDI_MI}, the waiting-time bottleneck
is removed when multiple QM banks ($k=10^4$) are available. The key rate
remains high even at long distances. Defining the effective transmission
distance as the point where the key rate drops to 90\% of its peak value, we obtain
approximately 169-173 km and 324-331 km for the protocols without and with a
repeater, respectively
(ranges reflect different 
multiplexing depths for $n=7,5,3,1$). The inclusion of a repeater at the midpoint
therefore nearly doubles the effective transmission distance compared
to direct transmission.

\subsection{MDI-QKD over a Repeater Chain}

The $n$-MUX MDI-QKD protocol extends naturally to a chain of untrusted repeaters, as outlined in Algorithm~\ref{alg:sdiqkd-multi-repeater}. Each elementary link generates $n$-MUX entanglement, which is connected via entanglement swapping to establish long-distance EPR pairs between Alice and Bob. 

This illustrates a key advantage of entanglement-based QKD: compatibility with quantum repeaters. In the MDI setting, only Alice and Bob’s devices are trusted, while all repeaters may be adversarial. Thus, the extension inherits the security guarantees of MDI-QKD while enabling scalable, long-distance secure communication with reduced memory requirements through multiplexing.

 While prepare-and-measure QKD protocols can be adapted to repeater architectures via entanglement swapping, such hybrids sacrifice the operational simplicity that originally motivated prepare-and-measure protocols and still require the complex quantum-memory hardware inherent to entanglement-based systems.

\begin{algorithm}[htbp]
\caption{MI $n$-MUX MDI-QKD over an $R$-repeater chain}
\label{alg:sdiqkd-multi-repeater}
\begin{algorithmic}[1]
\Require Trusted devices at Alice ($A$) and Bob ($B$); repeater chain $A=C_0,C_1,\dots,C_R,C_{R+1}=B$ with $R\ge1$ untrusted repeaters; time-bin depth $n$; authenticated classical channels; QMs at all nodes.

\State Alice and $C_1$, each intermediate pair of repeaters $(C_j,C_{j+1})$, and $C_R$ and Bob execute the $n$-MUX entanglement sharing protocol.  
In particular, Alice and Bob actively send their $n$-MUX photons toward $C_1$ and $C_R$, respectively, while all other nodes create $n+1$ shared EPR pairs across their adjacent links.

\State The repeater nodes perform entanglement swapping so that Alice and Bob obtain $n+1$ end-to-end EPR pairs.

\State Repeat Steps 1–2 until Alice and Bob share $2N$ end-to-end EPR pairs.

\State Alice and Bob use $N$ pairs for channel testing. On the remaining $N$ pairs, they apply an $(N,m)$ entanglement distillation protocol to obtain $m$ high-fidelity pairs.

\State Alice and Bob measure the distilled EPR pairs in the computational basis to obtain the secret key.
\end{algorithmic}
\end{algorithm}

\section{Discussion}\label{sec:con}

We have developed MA QKD protocols assisted by time-bin multiplexed photons, employing NV-center quantum memories, delay operators, and degree-of-freedom swap gates to enhance entanglement distribution.  Secret key rate analysis in lossy channels demonstrates improved rates and extended transmission distances compared to conventional protocols under limited bandwidth.

We further proposed a multiplexed-photon entanglement-based measurement-device-independent QKD protocol. By combining multiplexing with the MDI framework, the protocol improves robustness against channel losses and untrusted measurement devices at the relay node.

Overall, our results indicate that the proposed scheme provides a promising architectural framework toward long-distance quantum communication, establishing a feasible link between entanglement-based MDI-QKD and memory-assisted quantum repeater architectures.

 Exploring additional degrees of freedom, such as frequency multiplexing, may further enhance the capacity of such architectures and remains an interesting direction for future work.

\bibliographystyle{IEEEtran}
\bibliography{qkd}

\end{document}